\documentclass[modern]{aastex61}

\newcommand{\teff}{T$_{\rm eff}$}
\newcommand{\lbol}{$\log_{10}{\rm L/L_{bol}}$}
\newcommand{\lhalbol}{$\log_{10}{\rm L_{H\alpha}/L_{bol}}$}
\newcommand{\masyr}{mas\,yr$^{-1}$}
\newcommand{\kms}{km\,s$^{-1}$}
\newcommand{\age}{7.6$\pm$2.2}
\newcommand{\feh}{\lbrack Fe/H\rbrack}

\newcommand{\Msun}{M$_{\odot}$}

\received{1 June 2017}
\revised{TBD}
\accepted{TBD}
\submitjournal{ApJ}
\shorttitle{The Age of TRAPPIST-1}
\shortauthors{Burgasser \& Mamajek}

\begin{document}
\title{On the Age of the TRAPPIST-1 System}

\author[0000-0002-6523-9536]{Adam J.\ Burgasser} \affiliation{Department
  of Physics, University of California, San Diego, CA 92093, USA }
\&
\author[0000-0003-2008-1488]{Eric E. Mamajek}
\affiliation{Jet Propulsion Laboratory, California Institute of Technology,
  4800 Oak Grove Drive, Pasadena, CA 91109, USA}
\affiliation{Department of Physics \& Astronomy, University of Rochester,
  Rochester, NY 14627, USA}

\begin{abstract}
The nearby (d = 12 pc) M8 dwarf star TRAPPIST-1
(2MASS~J23062928$-$0502285) hosts a compact system of at least seven
exoplanets with sizes similar to Earth.
Given its importance for testing planet formation and evolution
theories, and for assessing the prospects for habitability among
Earth-size exoplanets orbiting the most common type of star in the
Galaxy, we present a comprehensive assessment of the age of this
system.
We collate empirical age constraints based on the color-absolute magnitude
diagram, average density, lithium absorption, surface gravity features,
metallicity, kinematics, rotation, and magnetic activity; and conclude that
TRAPPIST-1 is a transitional thin/thick disk star with an
age of {\age}~Gyr. 
The star's color-magnitude position is consistent with it being
slightly metal-rich ([Fe/H] $\simeq$ +0.06), in line
with its previously reported near-infrared spectroscopic metallicity; and it has a 
radius (R = 0.121$\pm$0.003~R$_{\odot}$) that is larger by 8--14\%
compared to solar-metallicity evolutionary models.
We discuss some implications of the old age of this system 
with regard to the stability and habitability of its planets.
\end{abstract}


\section{Introduction} \label{sec:intro}

TRAPPIST-1 (2MASS~J23062928$-$0502285; \citealt{2000AJ....120.1085G})
is as an ultracool M8 dwarf 12~pc from the Sun which was recently
identified to host at least seven Earth-sized planets, three orbiting
within the star's habitable zone
\citep{2016Natur.533..221G,2017Natur.542..456G,2017arXiv170304166L}.
The planets were identified by both ground-based and space-based
transit observations, and span orbit periods of 1.5--19~days and
orbital semi-major axes of 0.011-0.062~AU (21-114 stellar
radii). These observations and others have provided important
constraints on the physical parameters of the star itself, summarized
in Table~\ref{tab:star}.

One crucial parameter of TRAPPIST-1 that is poorly constrained by observations to date is
its age, due to the weak empirical age diagnostics currently available for ultracool M dwarfs.  
While rotation (gryochronology), activity diagnostics, and
lithium depletion are standard age-dating tools for solar-type stars
\citep{2010ARA&A..48..581S}, the physical properties 
of ultracool dwarfs restrict the application of these tools at the bottom
of the main sequence.  The low ionization of ultracool dwarf
photospheres reduces their coupling with magnetic winds,
resulting in spin-down timescales that can exceed the age of the
Milky Way Galaxy \citep{2008AJ....135..785W}. Depletion of lithium, which
provide approximate ages for solar-type stars up to 1--2~Gyr
\citep{2005A&A...442..615S}, is complete for fully-convective low-mass
stars by $\sim$200~Myr, and so is useful only for young ultracool dwarfs
\citep{1998ApJ...499L.199S}.  Spectral age diagnostics, such as
surface gravity sensitive features, are also limited to ages older
than $\sim$300~Myr \citep{2013ApJ...772...79A}.  While the kinematics
of TRAPPIST-1 suggest it to be an ``old disk'' star
\citep{1992ApJS...82..351L,2015ApJS..220...18B}, such labels are insufficient to firmly
constain an age.  \citet{2015ApJ...810..158F} report an fairly unconstrained age
constraint of 0.5--10~Gyr; while \citet{2017arXiv170304166L} adopt a more constrained,
but still broad, range of 3--8~Gyr. In contrast, several studies in the literature
have argued that TRAPPIST-1 must be young based on the strength of its nonthermal magnetic
emission \citep{2017A&A...599L...3B,2017MNRAS.469L..26O}. 

Age is necessary for understanding the formation, orbital
evolution, stability, and surface evolution (including habitability) of
the planets orbiting TRAPPIST-1. With this in mind, we present an analysis of various
age-related empirical diagnostics for this source that allow us to more precisely quantify
its age.  In Section~2 we analyze in turn the color-absolute magnitude diagram, average density,
lithium abundance, surface
gravity features, metallicity, kinematics, rotation, and magnetic
activity as age diagnostics for the star.  In Section~3 we summarize these into a concordance age of
{\age}~Gyr, and discuss implications on the stability and habitability of the planetary system.

\clearpage

\startlongtable
\begin{deluxetable}{lccl}
\tablecaption{Stellar Parameters for TRAPPIST-1 \label{tab:star}}
\tablehead{
\colhead{Parameter} & \colhead{Value} & \colhead{Units} & \colhead{Ref.}
}
\startdata
\hline
\multicolumn{4}{c}{Physical Parameters} \\
\hline
{\lbol}                 & $-$3.28$\pm$0.03   & dex & 1\\
{\teff}                 & 2560\,$\pm$\,50    & K   & 2\\
{}[Fe/H]                  & +0.04\,$\pm$\,0.08 & dex & 2\\
$P_{\rm rot}$              & 3.295$\pm$0.003   & day  & 3\\
$v\sin{i}$               & 6                 & \kms &  4\\
$\log_{10}$(L$_{H\alpha}$/L$_{bol}$) & $-$4.85 to $-$4.60    & dex & 5--8\\
$\log_{10}$(L$_{X}$/L$_{bol}$) & $-$3.52$\pm$0.17    & dex & 9 \\
Radius                   & 0.117$\pm$0.004     & R$_{\odot}$ & 2\\
\nodata                   & 0.121$\pm$0.003     & R$_{\odot}$ & 10 \\
Mass                     & 0.080$\pm$0.007   & M$_{\odot}$ & 2\\
Density                  & 50.7$^{+1.2}_{-2.2}$ & $\rho_{\odot}$ & 2\\
Age                      & {\age}            & Gyr & 10 \\
\hline
\multicolumn{4}{c}{Astrometric/Kinematic Parameters}\\
\hline
$\alpha$        &  346.6250957          & deg    & 11\\
$\delta$        &  -5.0428081 	        & deg    & 11\\
$\mu_{\alpha}$    &  922.0\,$\pm$\,0.6    & \masyr & 12 \\
$\mu_{\delta}$    & $-$471.9\,$\pm$\,0.9  & \masyr & 12 \\
$\pi$        & 80.09\,$\pm$\,1.17    & mas    & 12 \\
Distance        & 12.49\,$\pm$\,0.18    & pc    & 12 \\
$v_r$           & -51.688\,$\pm$\,0.014 & \kms  & 4\\
$U$             & -43.8\,$\pm$\,0.7     & \kms  & 10 \\
$V$             & -66.3\,$\pm$\,0.5     & \kms  & 10 \\
$W$             & 11.0\,$\pm$\,0.3      & \kms  & 10 \\
$S$             & 80.0\,$\pm$\,0.7      & \kms  & 10 \\
\hline
\multicolumn{4}{c}{Photometric Parameters}\\
\hline
$V$        & 18.75\,$\pm$\,0.03               & mag & 13\\
$R_{C}$     & 16.401\,$\pm$\,0.004 & mag & 14\\
$I_{C}$     & 13.966\,$\pm$\,0.002 & mag & 14\\
$J$        & 11.35\,$\pm$\,0.02   & mag & 15\\
$H$        & 10.72\,$\pm$\,0.02   & mag & 15\\
$K_s$      & 10.30\,$\pm$\,0.02   & mag & 15\\
$W1$       & 10.07\,$\pm$\,0.02   & mag & 11\\
$W2$       & 9.81\,$\pm$\,0.02    & mag & 11\\
$W3$       & 9.51\,$\pm$\,0.04    & mag & 11\\
$V-K_s$    & 8.45\,$\pm$\,0.04                & mag & 13,15\\
$M_V$      & 18.27\,$\pm$\,0.04   & mag & 10 \\
$M_J$      & 10.87\,$\pm$\,0.04   & mag & 10 \\
$M_{K_s}$   & 9.81\,$\pm$\,0.04    & mag & 10 
\enddata
\tablereferences{
(1) \citet{2015ApJ...810..158F};
(2) \citet{2016Natur.533..221G};
(3) \citet{2017arXiv170310130V};
(4) \citet{2014MNRAS.439.3094B};
(5) \citet{2000AJ....120.1085G}
(6) \citet{2008ApJ...684.1390R}
(7) \citet{2014MNRAS.439.3094B};
(8) \citet{2015ApJS..220...18B};
(9) \citet{2017MNRAS.465L..74W}
(10) This paper;
(11) AllWISE epoch 2010.5589 \citep{2013wise.rept....1C};
(12) \citet{2016AJ....152...24W}; 
(13) \citet{2015AJ....149....5W};
(14) \citet{2006PASP..118..659L};
(15) 2MASS \citep{2006AJ....131.1163S};
}
\end{deluxetable}

\section{Analysis}     

\subsection{Age Constraints from the HR Diagram}

\begin{figure}
\plottwo{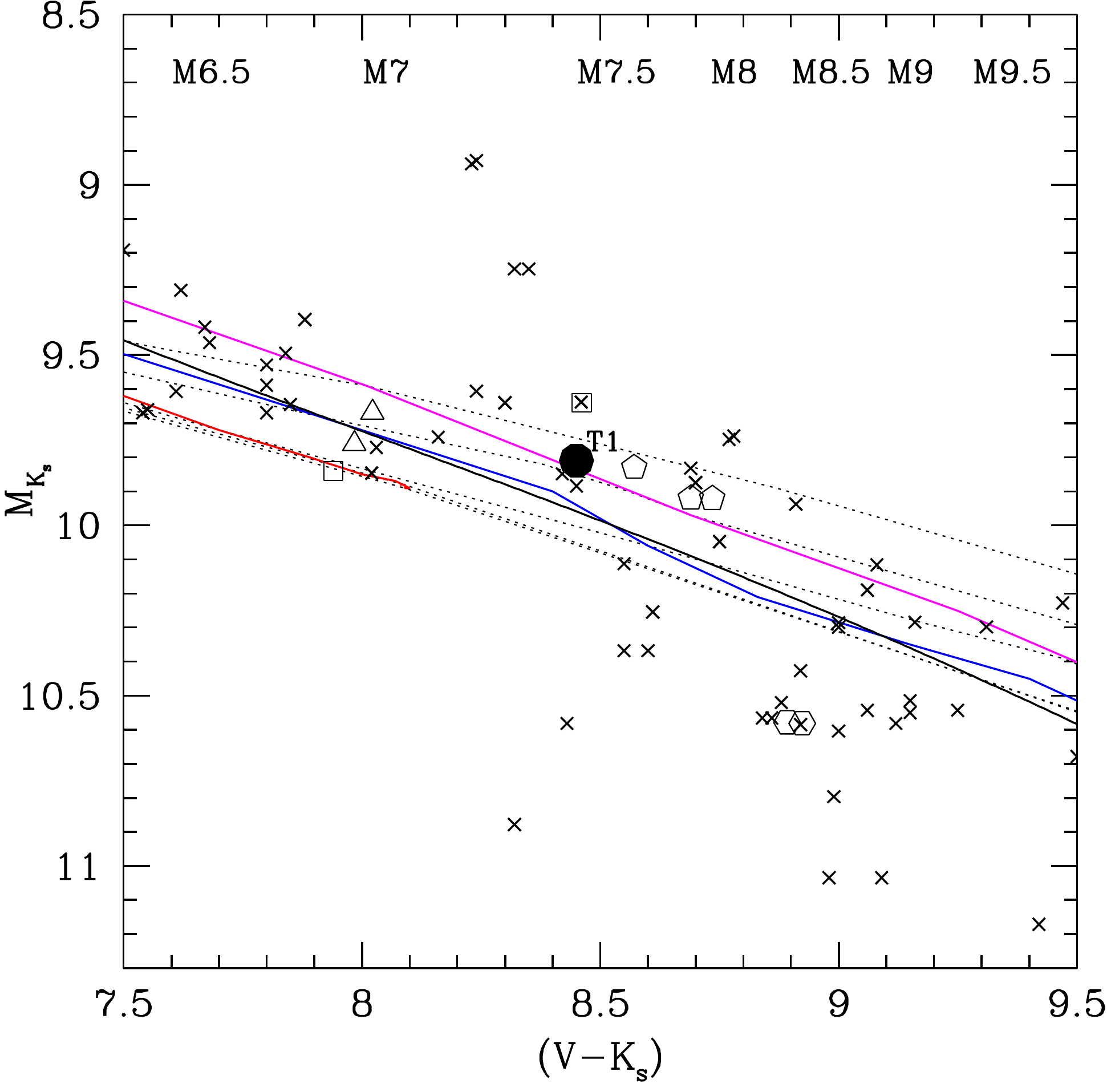}{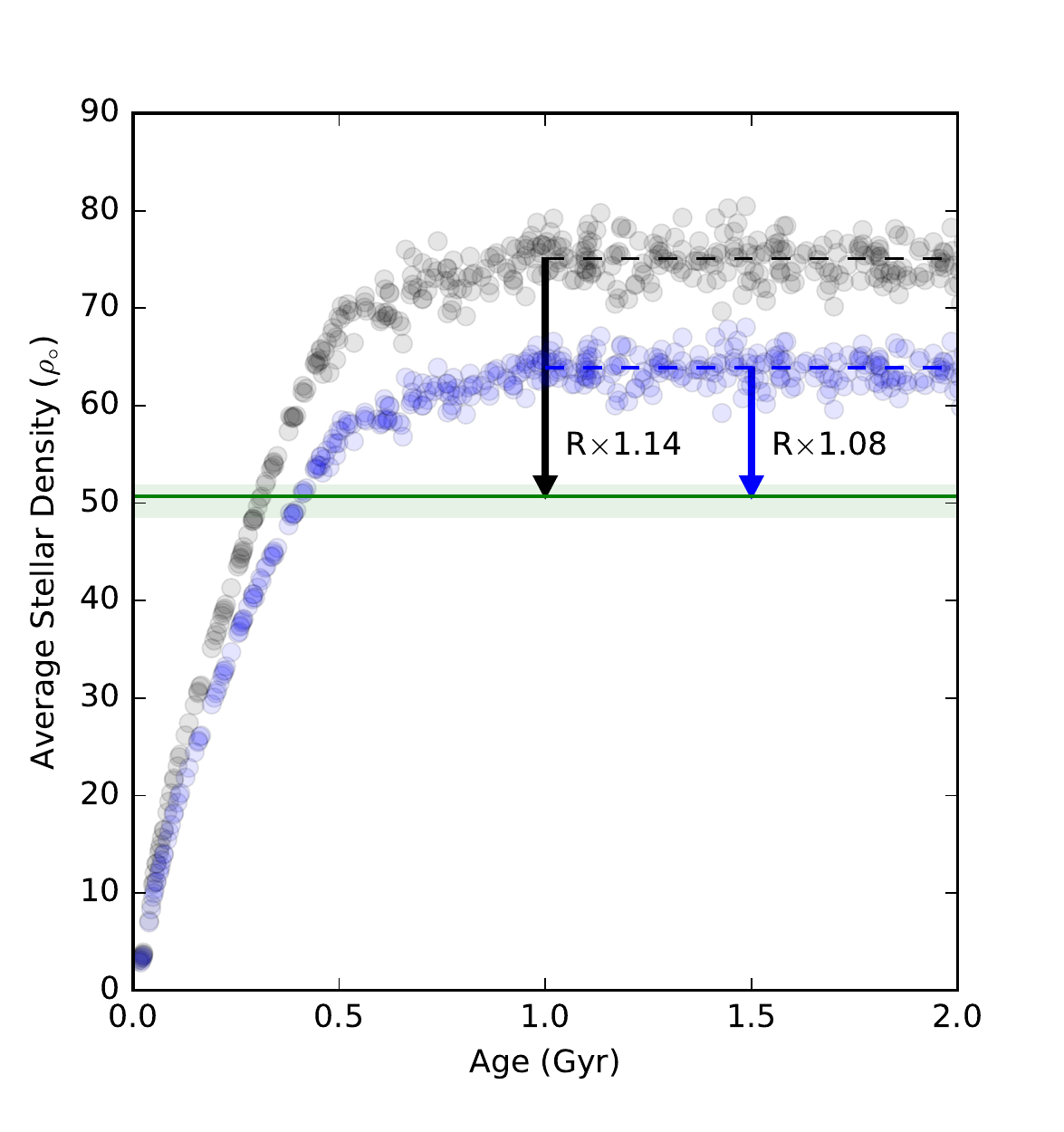}
\caption{(Left): Color-absolute magnitude diagram ($M_{Ks}$ versus $V-K_s$) for
  TRAPPIST-1, late-M spectral standard stars from \citet{1991ApJS...77..417K,1997ApJ...476..311K,2010ApJS..190..100K} and \citet{2002AJ....123.2002H,2004AJ....128.2460H}, and late-M field stars
  from \cite{2014AJ....147...94D} and \citet{2015AJ....149....5W}.  
  TRAPPIST-1 is
  plotted as a {\it large filled circle}; the spectral standards
  are indicated as {\it open triangles} (M7V), {\it open squares}
  (M7.5V), {\it open pentagons} (M8V), and {\it open hexagons}
  (M8.5V); all others as {\it crosses}. The color-magnitude locus for M dwarfs is
  plotted as a {\it solid black line}. 
   Isochrones from \citet{2015A&A...577A..42B} for ages of 0.2,
  0.3, 0.5, 3 and 8~Gyr ({\it dashed
    lines}, older isochrones toward the bottom) and masses of 0.07, 0.08, and 0.09 \Msun\
    (from top to bottom, {\it solid magenta}, {\it blue}, and {\it red} lines, respectively) are also shown.
    Given the slow evolution of M dwarfs on the
main sequence, the 3 and 8 Gyr isochrones are indistinguishable on
the scale of the diagram.
    Along the top are listed the approximate spectral type based on $V-K_s$ color.
(Right): Average stellar density in solar units as a function of age based on the theoretical models of
\citet[black symbols]{1997ApJ...491..856B,2001RvMP...73..719B} and \citet[blue symbols]{2015A&A...577A..42B}
for a luminosity {\lbol} = $-$3.28$\pm$0.03. Values were selected by Monte Carlo sampling in age (uniform distribution) and luminosity (normal distribution). The average density
measured for TRAPPIST-1 from \citet{2016Natur.533..221G}, 50.7$^{+1.2}_{-2.2}$~$\rho_{\odot}$, is indicated by the horizontal green line and region. The radius scale factors
needed to ``inflate'' the models for ages $>$1~Gyr (dashed lines) to the average density of TRAPPIST-1 are indicated by the arrows. 
    \label{fig:cmd}}
\end{figure}

The locations of TRAPPIST-1 and other late M-type spectral
standards and field stars on the $M_{Ks}$ versus ($V-K_s$) color-absolute
magnitude diagram (CMD) are shown in Figure~\ref{fig:cmd}. The comparison data 
are drawn from \citet{2014AJ....147...94D}
and \citet{2015AJ....149....5W} for nearby M dwarfs with trigonometric
parallaxes. A polynomial fit to these data between 3.5 $<$
($V-K_s$) $<$ 11.8 yields:
\begin{eqnarray}
\nonumber M_{Ks}\, =\, 26.987
- 20.6315(V-K_s)\,
+ 6.88044(V-K_s)^2\,
- 1.01665(V-K_s)^3\, \\
 + 0.0707374(V-K_s)^4\,
- 0.00188517(V-K_s)^5
\end{eqnarray}
\noindent The rms scatter among field M dwarfs is 0.48 mag. 
Recent spectroscopic surveys of field M dwarfs find
median metallicities ranging from \feh\, $\simeq$ +0.04 \citep{2014AJ....147...20N} to \feh\,
$\simeq$ -0.03 \citep{2015ApJ...804...64M}, with 1$\sigma$ dispersions of $\sim$0.2
dex. Hence, this CMD sequence is largely
representative of a solar metallicity population.\\

Also shown on the CMD are recent isochrones of low mass stars of
solar composition from \citet{2015A&A...577A..42B}
for ages of 0.2, 0.3, 0.5, 3, and 8~Gyr and masses of 
0.07, 0.08, and 0.09~M$_{\odot}$, spanning the
mass estimates for TRAPPIST-1. 
These isochrones reproduce the empirical locus for late-M dwarfs
to an accuracy of $\sim$0.1-0.2 mag in M$_{Ks}$ or $\sim$0.2 mag in ($V-K_s$) color;
but not the greater spread, which is likely due to metallicity
scatter \citep{2007ApJ...660..732L}. 
Binaries are relatively rare among late
M dwarfs ($\sim$10--20\%; \citealt{2007ApJ...668..492A,2012ApJ...757..141K})
and most field stars will be older than the pre-main-sequence contraction timescale of hundreds of Myr. 
  
TRAPPIST-1 lies 0.15$\pm$0.04~mag above (brighter than) the empirical locus.
This offset could be interpretted as TRAPPIST-1 being
a 0.2--0.3 Gyr-old, 0.07 \Msun\, brown dwarf.
However, this interpretation would predict a
density of 38~$\rho_{\odot}$ based on the \citet{2015A&A...577A..42B}
models, which is much lower than the
measured density from transit observations (see below).
We can also rule out contamination from a stellar or substellar
companion based on 
previous adaptive optics surveys \citep{2003ApJ...598.1265S,2003AJ....126.1526B,2003AJ....125.3302G,2012ApJ...754...44J} and multi-epoch radial velocity surveys \citep{2012ApJS..203...10T,2014MNRAS.439.3094B}.

Instead, the magnitude offset of TRAPPIST-1 is likely due to a 
slightly super-solar metallicity.  
The offset is similar to that of the M8~V standard VB~10
(aka GJ~752B), and it and its M3~V companion GJ~752A both
have \feh\, $\simeq$ 0.1 \citep{2014AJ....147...20N,2015ApJ...804...64M}, consistent with the
spectroscopic value for TRAPPIST-1 
\citep{2016Natur.533..221G}.
Alternately, attributing the 0.48 mag scatter in absolute magnitude for the CMD empirical 
locus entirely to metallicity scatter ($\sim$0.2 dex) would imply a metallicity-absolute
magnitude gradient of $\Delta$[Fe/H]/$\Delta$M$_{Ks}$ $\simeq$ $-$0.4
dex\,mag$^{-1}$, with metal-rich stars being brighter.
Under this interpretation, the $-$0.15~mag absolute magnitude offset for 
TRAPPIST-1 would be consistent with [Fe/H] = +0.06~dex, again consistent
with the spectroscopic metallicity.

Unfortunately, currently-available evolutionary models for the very lowest mass
stars and brown dwarfs are defined only for solar metallicities, preventing us from disentangling age
and metallicity effects. Our
CMD analysis therefore reinforces a supersolar metallicity
for TRAPPIST-1, but cannot constrain its age.\\

\subsection{Age Constraints from Stellar Density}

Average stellar density,  an observable from transit lightcurve analysis \citep{2003ApJ...585.1038S}, 
provides an independent check on the evolutionary state of TRAPPIST-1.
Figure~\ref{fig:cmd} shows the 
predicted stellar densities for stars and brown dwarfs from the solar-metallicity evolutionary models of \citet{1997ApJ...491..856B,2001RvMP...73..719B} and \citet{2015A&A...577A..42B} as a function of age, constrained to have the
observed luminosity of TRAPPIST-1. At this luminosity, young, contracting, substellar objects have densities that increase up to an age of approximately 500~Myr; beyond 1~Gyr (corresponding to hydrogen-burning low mass stars), densities plateau. The Burrows et al.\ models predict densities 17\% higher than the Baraffe et al.\ models, corresponding to radii that are 6\% smaller. 
The average density of TRAPPIST-1 is on the rising portion of this trend, and intersects with the Burrows et al.\ and Baraffe et al.\ models
at ages of 0.3~Gyr and 0.4~Gyr, respectively, corresponding to masses at the hydrogen-burning mass limit (0.071~M$_{\odot}$ and 0.079~M$_{\odot}$). 

These comparisons again suggest that TRAPPIST-1 is relatively young.
However, two factors must be considered. First, there are the previously-noted metallicity effects.
Among $-$0.04 $<$ [Fe/H] $<$ 0.12 low-mass stars,
\citet{2007ApJ...660..732L} found evolutionary models underpredict stellar radii by 10--20\%, a correction factor sufficient to bring the density plateaus of the Baraffe et al.\ and Burrows et al.\ models in line with the observed density of TRAPPIST-1.
Second, magnetic activity can also modulate the radii of low mass stars and brown dwarfs
\citep{2007ApJ...660..732L,2007ApJ...671L.149R,2007A&A...472L..17C,2009ApJ...700..387M,2010ApJ...722.1138M}.
Empirical relations from \citet{2012ApJ...756...47S} predict a modest effect: 
0--4\% based on X-ray emission, 4--6\% based on H$\alpha$ emission. However,
theoretical models by \citet{2007A&A...472L..17C} show radii can range over 0.10--0.14~R$_{\odot}$ at M = 0.08~M$_{\odot}$
for (black) spot coverage of up to 50\%, more than sufficient to cover the offset in TRAPPIST-1's stellar density.
{\em Kepler} data have confirm the presence of cool, stable magnetic spots on TRAPPIST-1 \citep{2017arXiv170304166L,2017arXiv170310130V}, so these may play a role in radius inflation.

Given that both metallicity and magnetic activity likely play a role in setting the average stellar density of TRAPPIST-1, the current unavailability of appropriate evolutionary models encapsulating these effects 
again prevents us from extracting meaningful age constraints from this physical parameter.

\subsection{Age Constraints from Lithium Depletion}

Multiple studies have reported the absence of 6708~{\AA} Li~I
absorption in the optical spectrum of TRAPPIST-1, indicating depletion
of this element in its fully convective interior
\citep{2009ApJ...705.1416R,2015ApJS..220...18B}. As noted above,
theoretical models of solar-metallicity stars and brown dwarfs
more massive than 0.06~M$_{\odot}$ show full depletion within $\sim$200~Myr
\citep{1997ApJ...482..442B,2004ApJ...604..272B}. Similarly, the \citet{1997ApJ...491..856B,2001RvMP...73..719B} and \citet{2015A&A...577A..42B}
evolutionary models predict a minimum age of 190~Myr for a
0.06~M$_{\odot}$ brown dwarf with the observed luminosity of
TRAPPIST-1. These
estimates provide a lower limit to TRAPPIST-1's age that is already
incorporated into the age range proposed by
\citet{2015ApJ...810..158F}.

\subsection{Age Constraints from Surface Gravity Features}

\begin{figure}
\plottwo{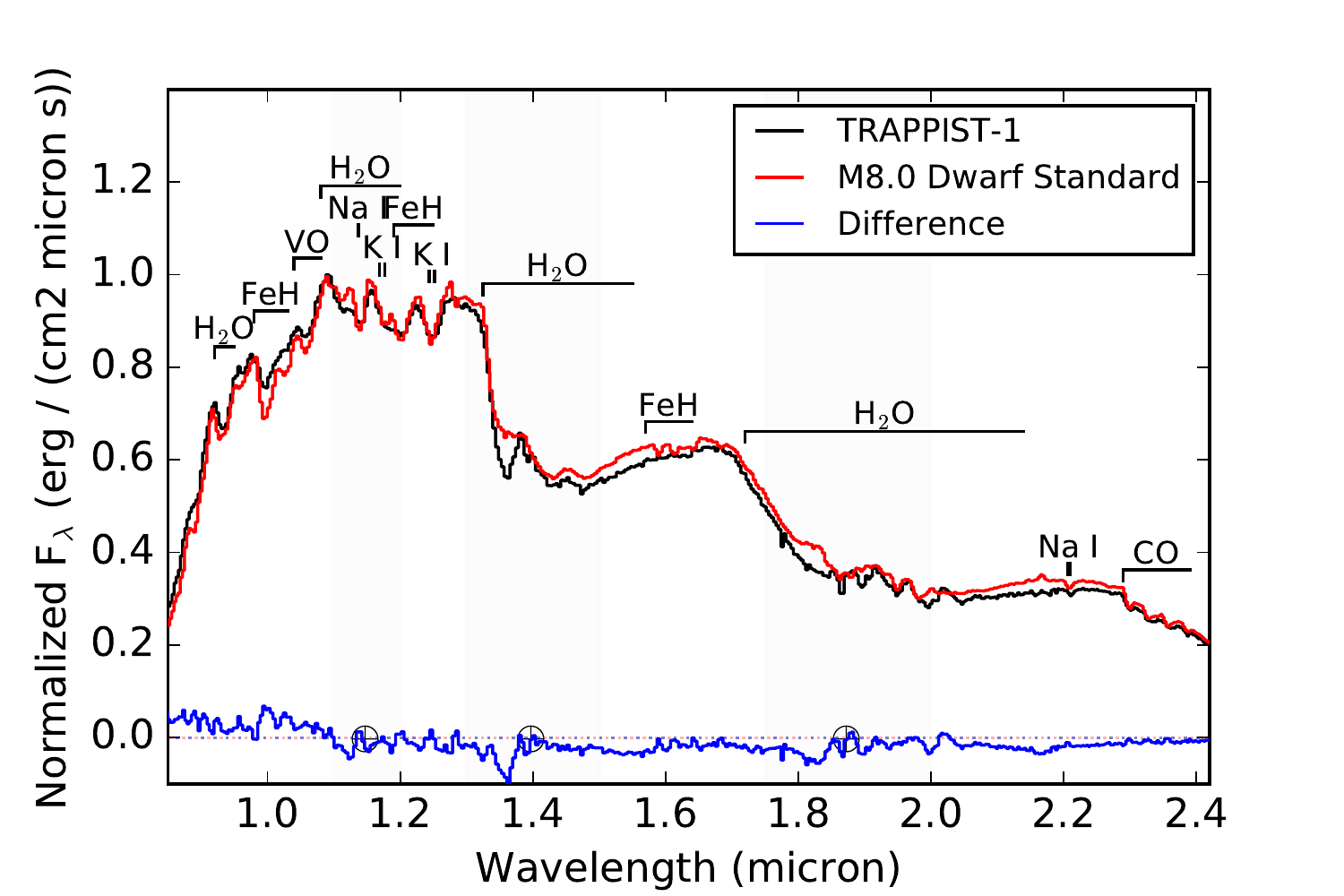}{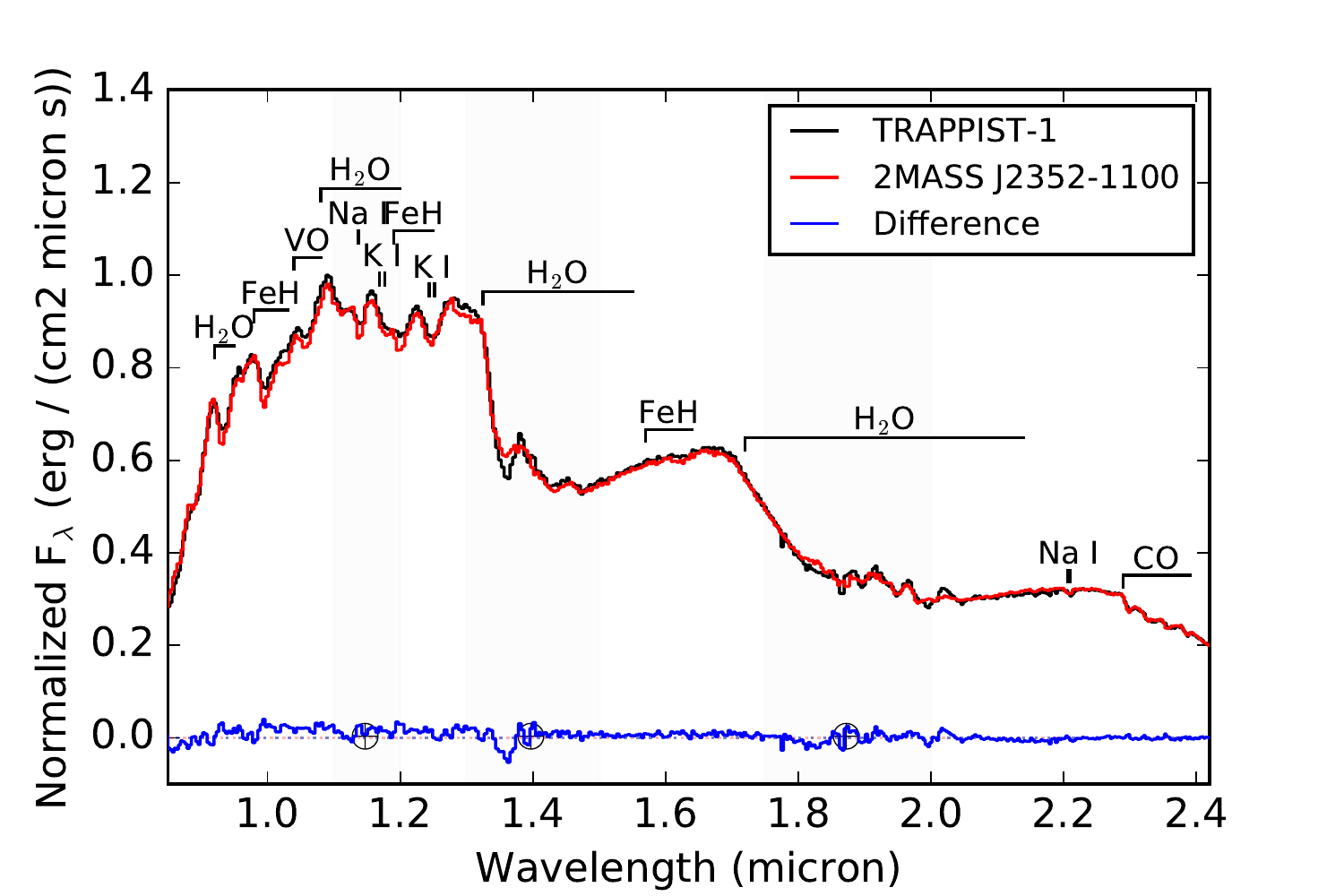} \\
\plottwo{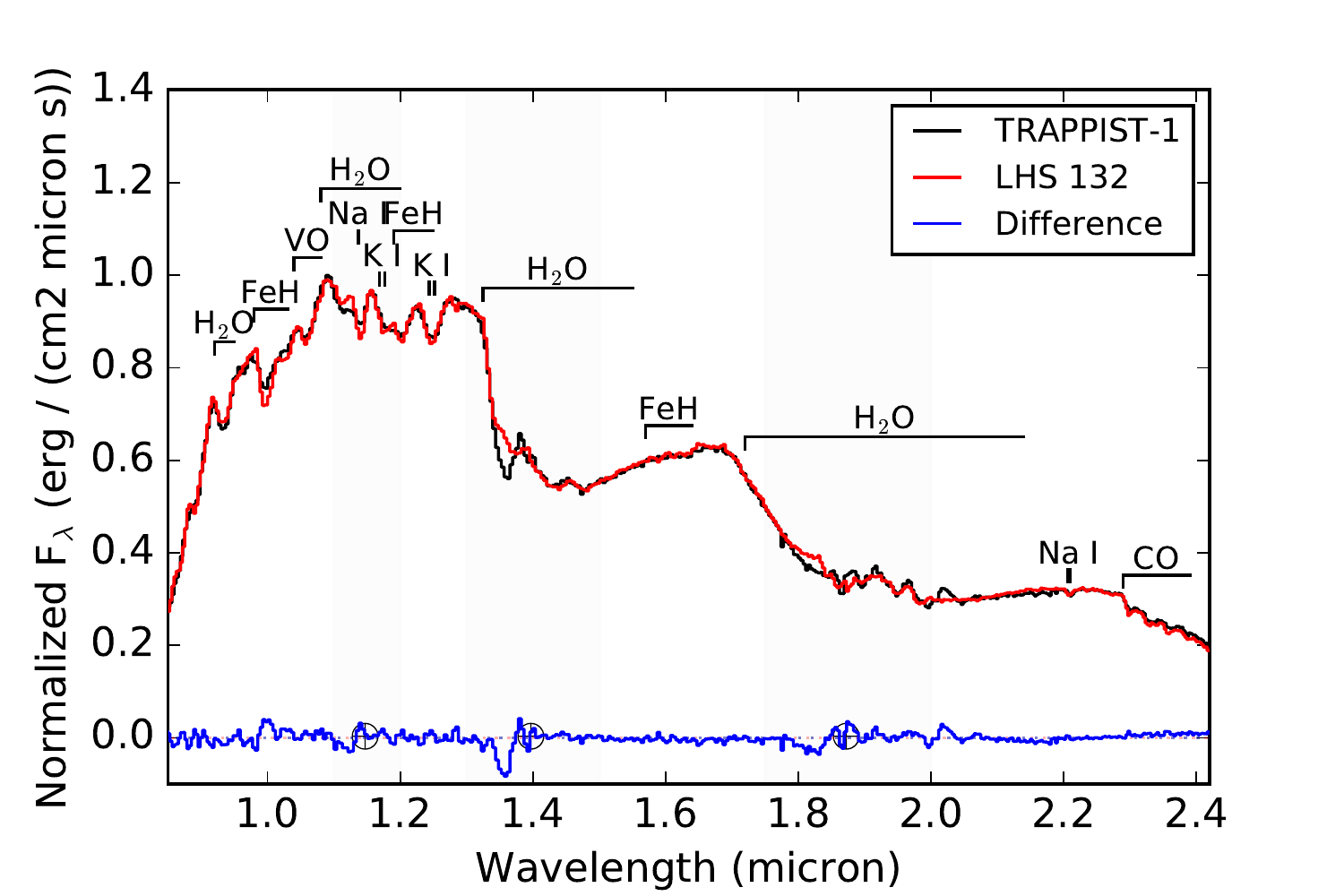}{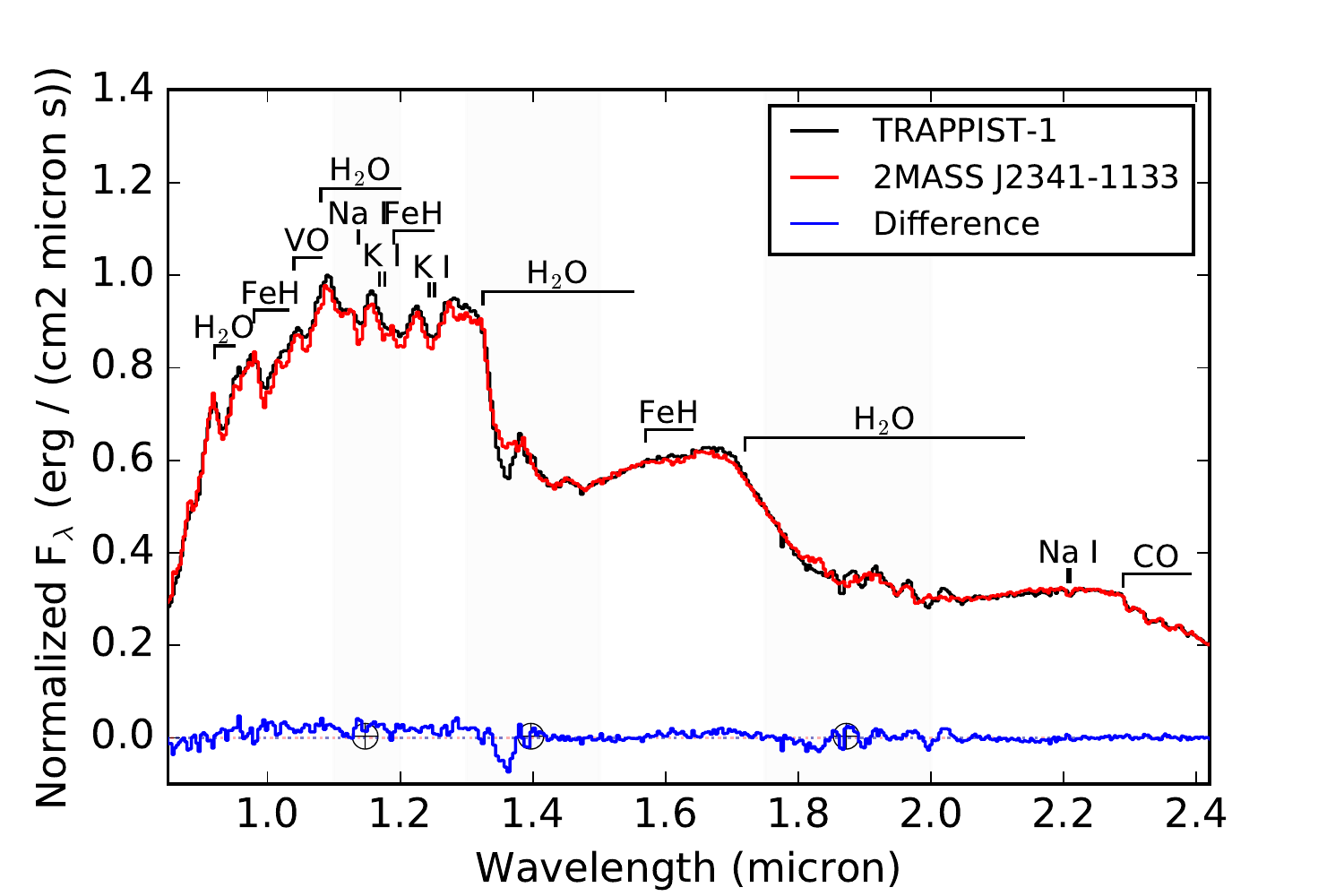}
\caption{Comparison of the low-resolution near-infrared spectrum of
  TRAPPIST-1 (\citealt{2016Natur.533..221G}; black line) to equivalent
  data (red lines) for the M8 standard VB~10 (top left) and the M8 dwarfs
  2MASS~J2352050$-$110043 (top right), LHS~132 (bottom left),
  and 2MASS~J2341286-113335 (bottom right). All comparison spectra are from
  \citet{2014ApJ...794..143B}, and are normalized to
  optimize agreement to the spectrum of TRAPPIST-1 outside telluric absorption bands (grey
  regions). Difference spectra are shown in blue. Absorption features
  attributable to Na~I, K~I, H$_2$O, CO, VO and FeH are labeled.
\label{figure:comp2352}}
\end{figure}

Low-resolution near-infrared spectra presented in
\citet{2016Natur.533..221G} are generally consistent with the M8 dwarf
spectral standard VB~10. Howeer, there are some notable peculiarities,
including weaker FeH absorption and a more triangular $H$-band peak
(Figure~\ref{figure:comp2352}). Application of the
\citet{2013ApJ...772...79A} surface gravity indices yields a gravity
classification of INT-G for this source, suggesting a low surface
gravity and young ($\sim$100-300~Myr) age.  Moreover, comparison of its near-infrared
spectrum to the entirety of the SpeX Prism Library
\citep{2014ASInC..11....7B} uncovers an excellent match to the M7
dwarf 2MASS~J2352050$-$110043 (\citealt{2007AJ....133..439C};
hereafter 2MASS~J2352$-$1100), a source identified by
\citet{2015ApJ...798...73G} and \citet{2016ApJ...821..120A} as a
possible kinematic member of the $\sim$110~Myr AB Doradus association
\citep{2005ApJ...628L..69L,2013ApJ...766....6B}. These lines of evidence again 
suggest TRAPPIST-1 could be a young brown dwarf.

However, 2MASS~J2352$-$1100 lacks
Li~I absorption, expected for a 110~Myr M8 dwarf; and it kinematic
association with AB Doradus is based on proper motion and estimated
distance alone, and may be spurious. TRAPPIST-1's kinematics firmly
rule out membership in AB~Doradus or any nearby young moving group
\citep{2013ApJ...762...88M,2014ApJ...783..121G}; and neither
TRAPPIST-1 nor 2MASS~J2352$-$1100 exhibit the enhanced VO absorption
generally seen in the spectra of low-gravity M and L dwarfs \citep{2008ApJ...689.1295K,2009AJ....137.3345C,2013ApJ...772...79A}.
The near-infrared spectrum of TRAPPIST-1 is also a good match to those of the M8 dwarfs LP~938-71
and 2MASS~J2341286-113335 (Figure~\ref{figure:comp2352}), neither of which are reported to be unusually young or active
\citep{2007AJ....133..439C,2007AJ....133.2258S}.

We conclude that the
INT-G gravity classifications for both sources are unrelated to youth,
and may arise from other physical factors, as previously reported for the high
velocity M6.5 2MASS~J02530084+1652532 (aka Teegarden's star, $V_{tan}$
= 93~{\kms};
\citealt{2003ApJ...589L..51T,2006AJ....132.2360H,2015ApJS..219...33G})
and the d/sdM7 metal-poor companion GJ~660.1B
\citep{2016AJ....151...46A}.
Given the apparent interplay between surface gravity and metallicity 
effects in index-based gravity metrics for late M dwarfs, we discount the INT-G
classification as evidence of youth for TRAPPIST-1.

\subsection{Age Constraints from Metallicity}

While age-metallicity correlations are generally weak among stellar
populations, the dispersion of stellar metallicities increases for populations older than a few
Gyr, and median metallicity decreases for populations older than
$\approx$10~Gyr \citep{1993A&A...275..101E, 2013A&A...560A.109H,
  2014A&A...565A..89B}. Hence, comparison of TRAPPIST-1's metallicity
to population distributions can provide a statistical constraint on its age.

\begin{figure}
\plottwo{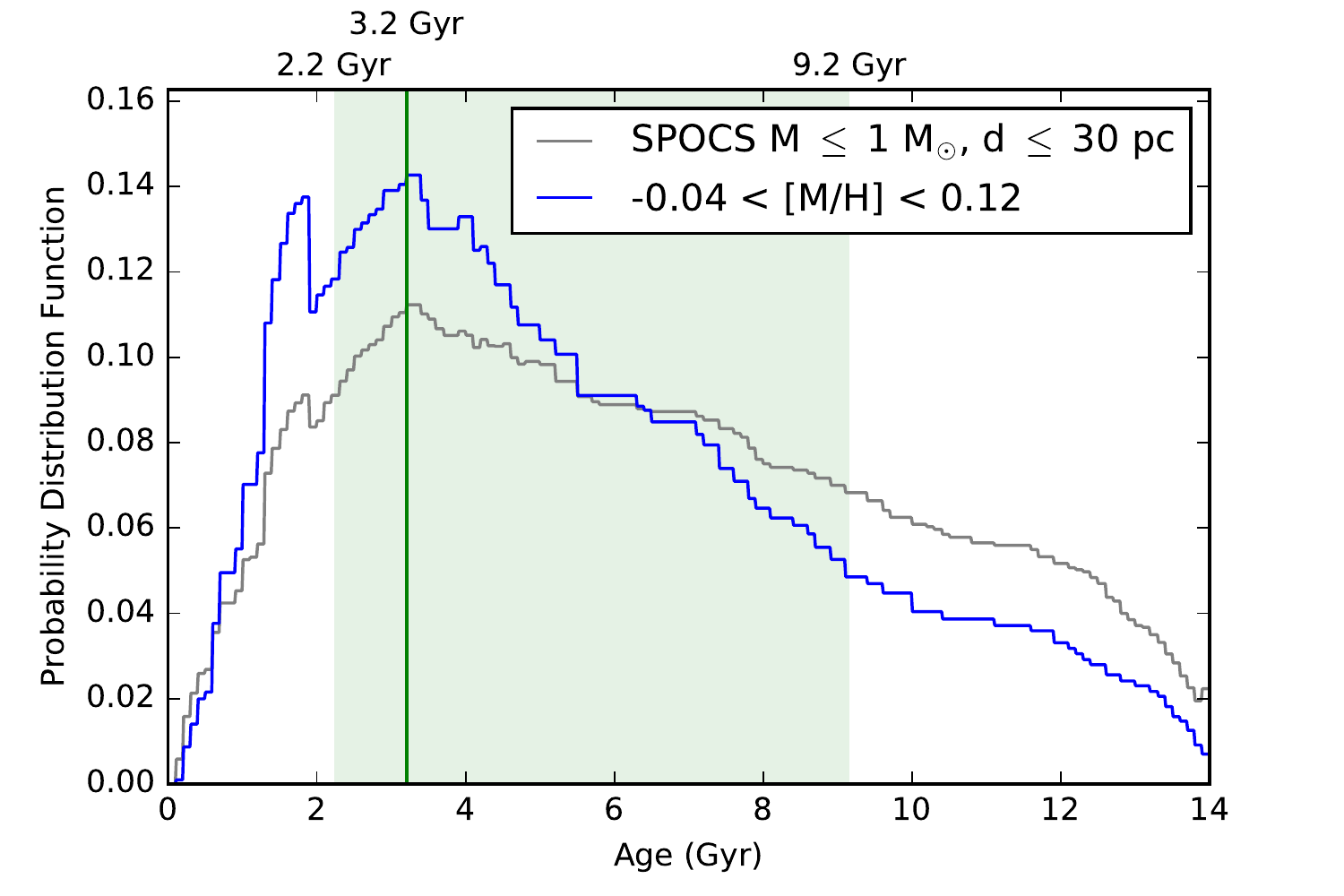}{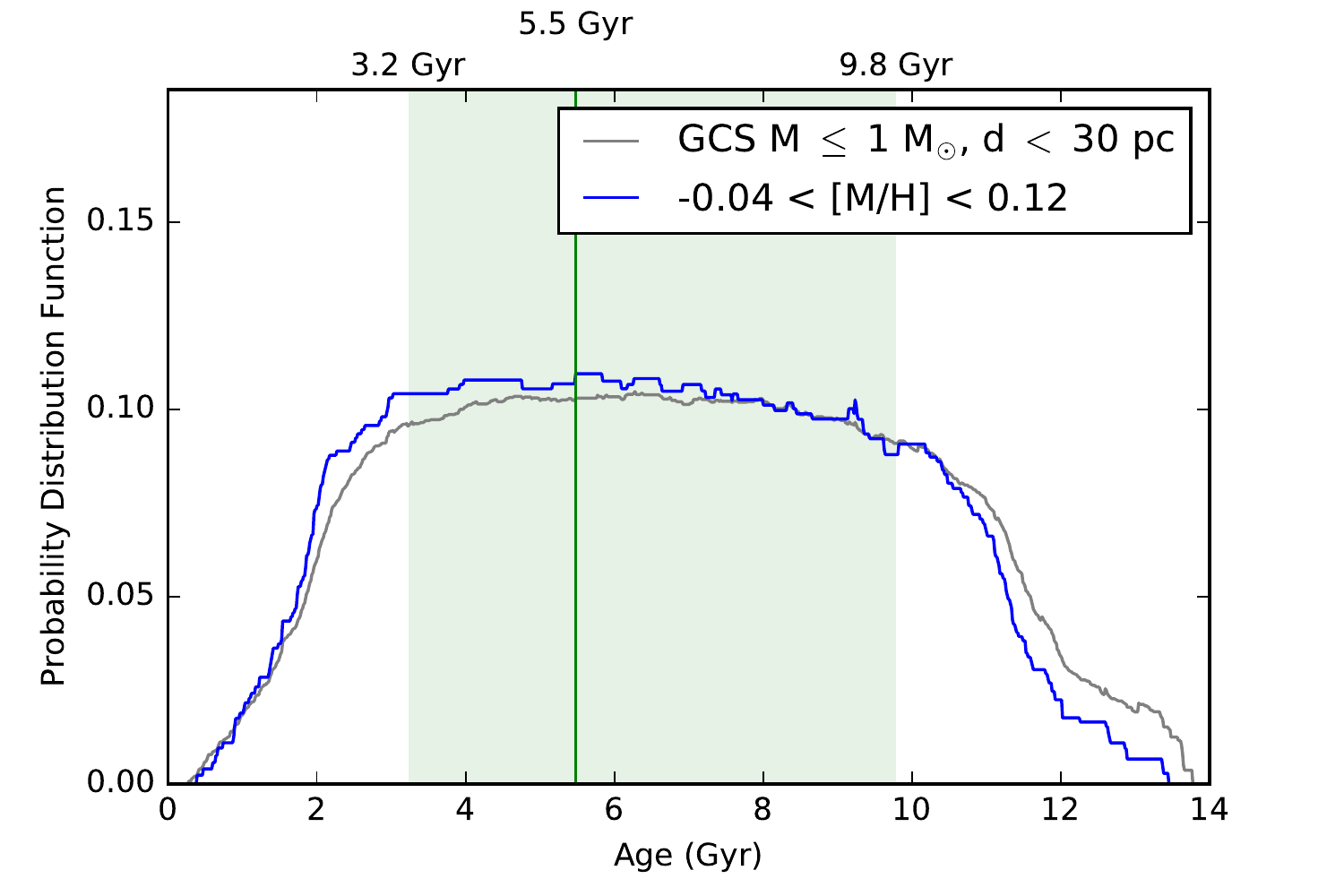}
\caption{Age probability distribution functions for stars with M $\leq$
  1~M$_{\odot}$, $-$0.04 $<$ [Fe/H] $<$ +0.12 and distances $\leq$
  30~pc in the SPOCS (left) and GCS (right) catalogs. Grey histograms
  are without the metallicity constraint; blue histograms are with the
  metallicity constraint. The solid vertical lines indicate the maximum likelihoods
  of the age distributions with all constraints
  and the shaded green regions encompass the 16\%
  to 84\% probability ranges.
\label{figure:metallicity}}
\end{figure}

To quantify this diagnostic, we examined the age and metallicity
distributions of stars drawn from the Spectroscopic Properties of Cool
Stars (SPOCS; \citealt{2005ApJS..159..141V}) and an updated analysis
of the Geneva-Copenhagen Survey (GCS;
\citealt{2011A&A...530A.138C}). In both samples, ages are inferred
by comparison of CMDs to model isochrones (for GCS, we used the ages
inferred from the Padova isochrones; \citealt{2008A&A...484..815B}),
while metallicities are determined from spectroscopic measurements in
SPOCS and Str\"omgren photometry in GCS.  For both samples, we selected M
$\leq$ 1~M$_{\odot}$ stars with $-$0.04 $<$ \feh\, $<$ +0.12 and
parallactic distances within 30~pc, and constructed an age
distribution by assuming a uniform likelihood for each star between the
minimum and maxium ages (SPOCS) or 16\% and 84\% Padova isochronal
ages (GCS).  These distributions are shown in
Figure~\ref{figure:metallicity}. In both samples, stars younger than 1--2~Gyr and older than
11--12~Gyr are relatively rare. The SPOCS age distribution peaks at young ages,
which is enhanced by the metallicity constraint. The GCS age
distribution is flat between 2--10~Gyr with a slight preferance toward
younger ages with the metallicity constraint. While the maximum
likelihood ages are quite different between these samples, their
 distributions overlap, and we infer ages of
3.2$^{+6.0}_{-1.0}$~Gyr for SPOCS and 5.5$^{+3.7}_{-2.3}$~Gyr for GCS.
As anticipated, the uncertainties are considerable and do little to
reduce the overall uncertainty in the system's age.

\subsection{Age Constraints from Kinematics}

\citet{2009ApJ...705.1416R} and \citet{2015ApJS..220...18B} previously
reported equivalent $UVW$ kinematics for TRAPPIST-1 based on radial
velocity and proper motion measurements, the latter study
concluding that the star is a borderline thin/thick disk star based on the
criteria of \citet{2003A&A...410..527B}.  We update this analysis
using the more precise radial velocity reported in
\citet{2014MNRAS.439.3094B} and astrometry reported in
\citet{2016AJ....152...24W}. The corresponding heliocentric $UVW$
velocities are given in Table~\ref{tab:star}.  Adopting the Local
Standard of Rest (LSR) correction of \citet{1998MNRAS.298..387D} used
by \citet{2003A&A...410..527B}, $UVW_{\odot}$ = [+10.00,+5.25,+7.17],
we find probabilities of kinematic association $P$(thin) = 81\%,
$P$(thick) = 19\% and $P$(halo) $<$ 0.1\%. With $P$(thick)/$P$(thin) =
0.23, this star remains a borderline thin/thick disk star by the
\citet{2003A&A...410..527B} criteria.

\begin{figure}
\plottwo{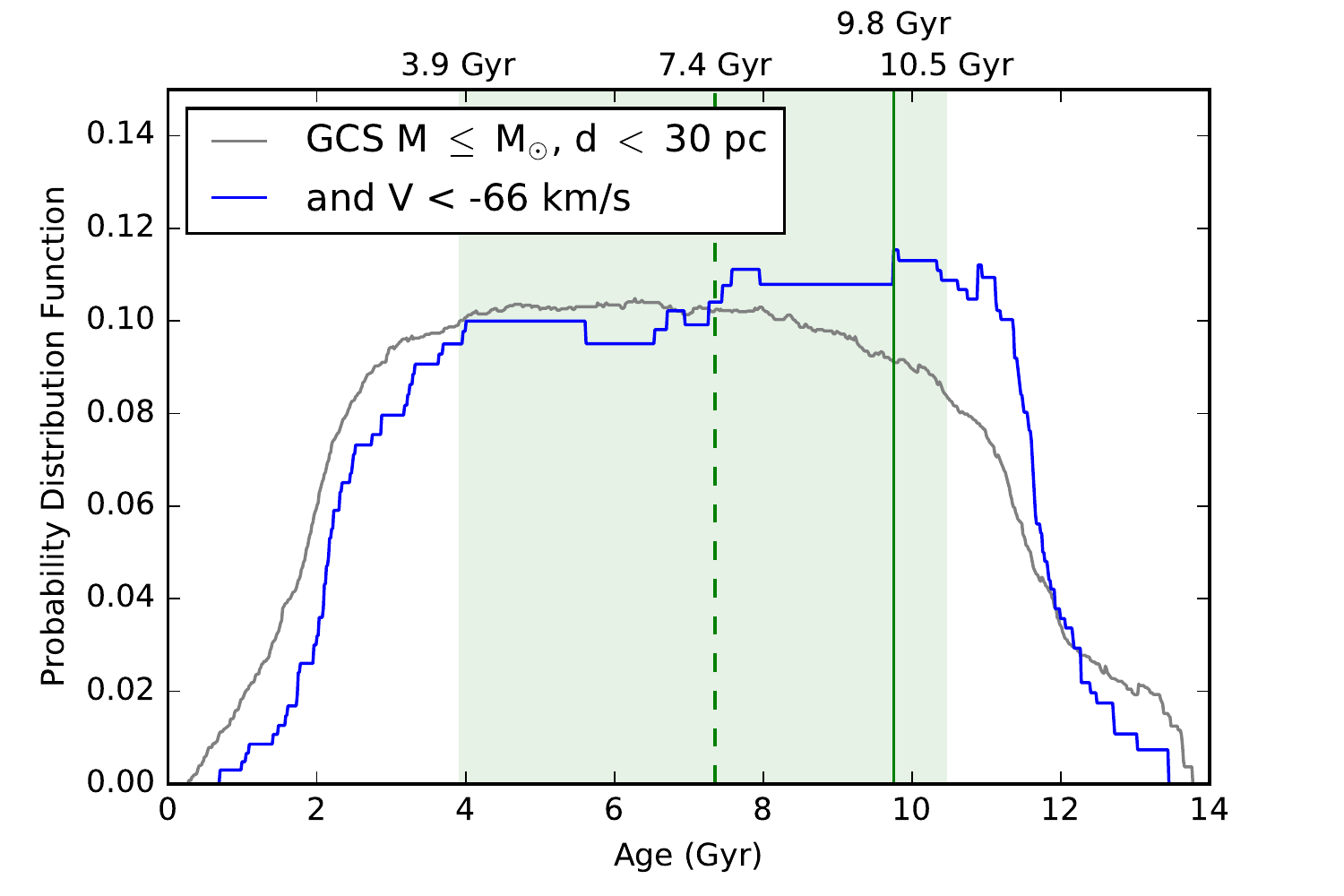}{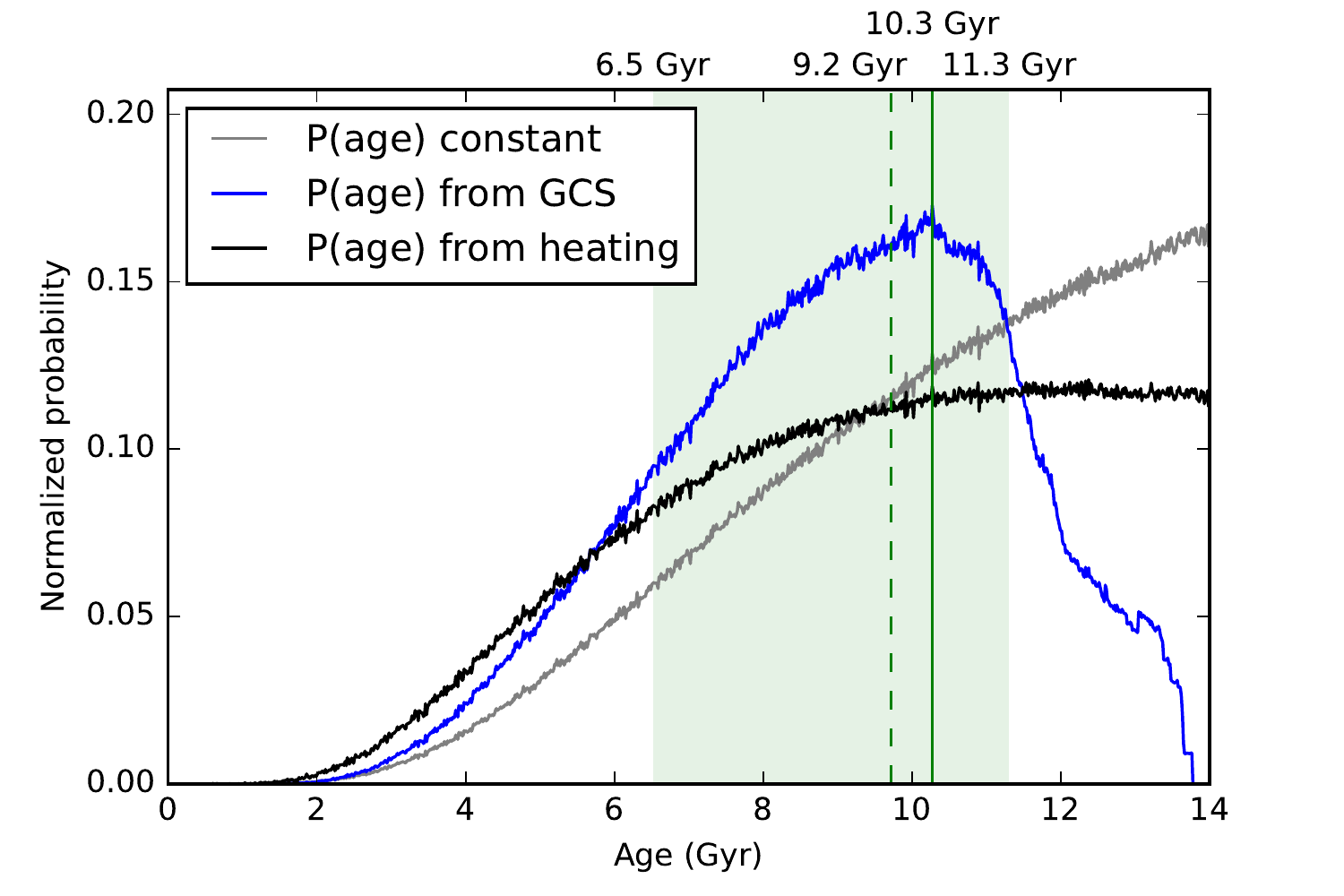}
\caption{(Left) Distribution of ages of M $<$ 1~M$_{\odot}$ GCS stars
  (grey) compared to those with $V$ $<$ $-$66~{\kms}
  (blue). (Right) Distribution of ages for a simulated population with
  $UVW$ dispersions based on three different age priors: a constant
  prior (grey), a prior based on the GCS sample (M $\leq$
  1~M$_{\odot}$ and $d \leq$ 30~pc; blue) and a prior based on a model
  for kinematic heating (black).  In both panels, the solid vertical
  lines indicate the maximum likelihood values of the age
  distributions for the velocity-selected (left) and GCS prior (right)
  samples, the dashed vertical lines the medians, and the shaded green
  regions encompass the 16\% to 84\% probability ranges.
\label{figure:kinematics} }
\end{figure}

To derive a quantitative estimate of TRAPPIST-1's kinematic
age, we took advantage of the fact that the $V$-velocity asymmetric drift
of stellar populations increases over time as the velocity scatter
increases ($V_a \propto \sigma_U^2$;
\citealt{1924ApJ....59..228S}). We again used the GCS sample and
examined the age distribution of stars with negative $V$
velocities like TRAPPIST-1. Figure~\ref{figure:kinematics} shows the same mass- and
distance- constrained distribution as Figure~\ref{figure:metallicity},
but with the additional constraint that $V$ $<$ $V_{T1}$ =
$-$66~{\kms}.  In this case we see a tilt toward older ages, with a
maximum likelihood value of 9.8~Gyr, albeit with a wide
uncertainty range (3.9--10.5~Gyr).  We find a similar distribution
(albeit with a very small sample) when a metallicity constraint was
also applied to the GCS sample.

As a second approach, we applied a Bayesian method to solve for the
probability distribution function of a star's age given its UVW
velocities:
\begin{equation}
P({\rm age|UVW}) \propto P({\rm UVW|age})P({\rm age})
\end{equation}
Here, $P$(age) is the {\it apriori} distribution of stellar ages while
$P$(UVW$|$age) is the distribution of stellar UVW velocities as they
evolve over time.  We considered three different age priors in our
analysis: a constant age distribution (constant star
formation rate) up to 12~Gyr; the age distribution of
GCS stars with M $\leq$ 1~M$_{\odot}$ and $d \leq$ 30~pc
without constraints on metallicity or $V$-velocity; and a constant age distribution with an additional ``heating''
term modeled after \citet{2009MNRAS.397.1286A} that deweights older
populations that spend less time in the immediate Solar Neighborhood,
\begin{equation}
\ln{P} \propto -\frac{\Delta\Phi(Z)}{\sigma_W(t)^2}
\end{equation}
Here, $\Delta\Phi(Z)$ is the vertical gravitational potential
difference from the mid-plane to a Galactic height $Z$, taken to
be 50~pc; and $\sigma_W(t)$ is the vertical velocity dispersion in
{\kms} over time $t$ in Gyr, calculated as
\begin{equation}
\sigma_W(t) = 8.388(t+0.01)^{0.445}
\end{equation}
\citep{2009MNRAS.397.1286A}. For the Galactic potential, we used the
four-component disk, halo, and bulge model of
\citet{2016A&A...593A.108B} at a Galactic radius of 8~kpc.  UVW
velocities as a function of age were drawn using the age-dispersion
relations of \citet{2009MNRAS.397.1286A}, including an asymmetric
drift term $V_a = -\frac{\sigma_U^2}{74~{\rm km~s^{-1}}}$. We drew 10$^7$ stars
at each simulated time step, and computed the fraction of draws as a
function of age whose UVW velocities were within 5$\sigma$ of
those of TRAPPIST-1.  

Figure~\ref{figure:kinematics} shows the
resulting distribution of stellar ages for our three age priors. All are strongly skewed toward older ages, with maximum
likelihood values ranging from 10.2~Gyr (GCS) to 12~Gyr
(constant). The GCS and heating model priors produce similar
distributions within 11~Gyr, the former dropping off rapidly beyond
this in accordance with the underlying sample age distribution.  Using
the median values as best estimates, we infer ages of
9.2$^{+1.1}_{-2.7}$~Gyr and 8.7$^{+2.3}_{-2.9}$~Gyr for the GCS and
heating model priors, which are on the high end but consistent with
the other age diagnostics.

\subsection{Age Constraints from Rotation}

While the timescales for rotation spindown and activity decline for ultracool
dwarfs become exceedingly long compared to more massive stars
\citep{2008AJ....135..785W,2011ApJ...727...56I}, there is evidence
that both properties do evolve in a measureable way (e.g.,
\citealt{2015ApJS..220...18B}). Early analysis of TRAPPIST-1 has
suggested that its 3.295~day rotation period is average for late-type M dwarfs, suggesting a 
``middle-age'' star (3--8~Gyr from \citealt{2017arXiv170304166L}).

\begin{figure}
\plottwo{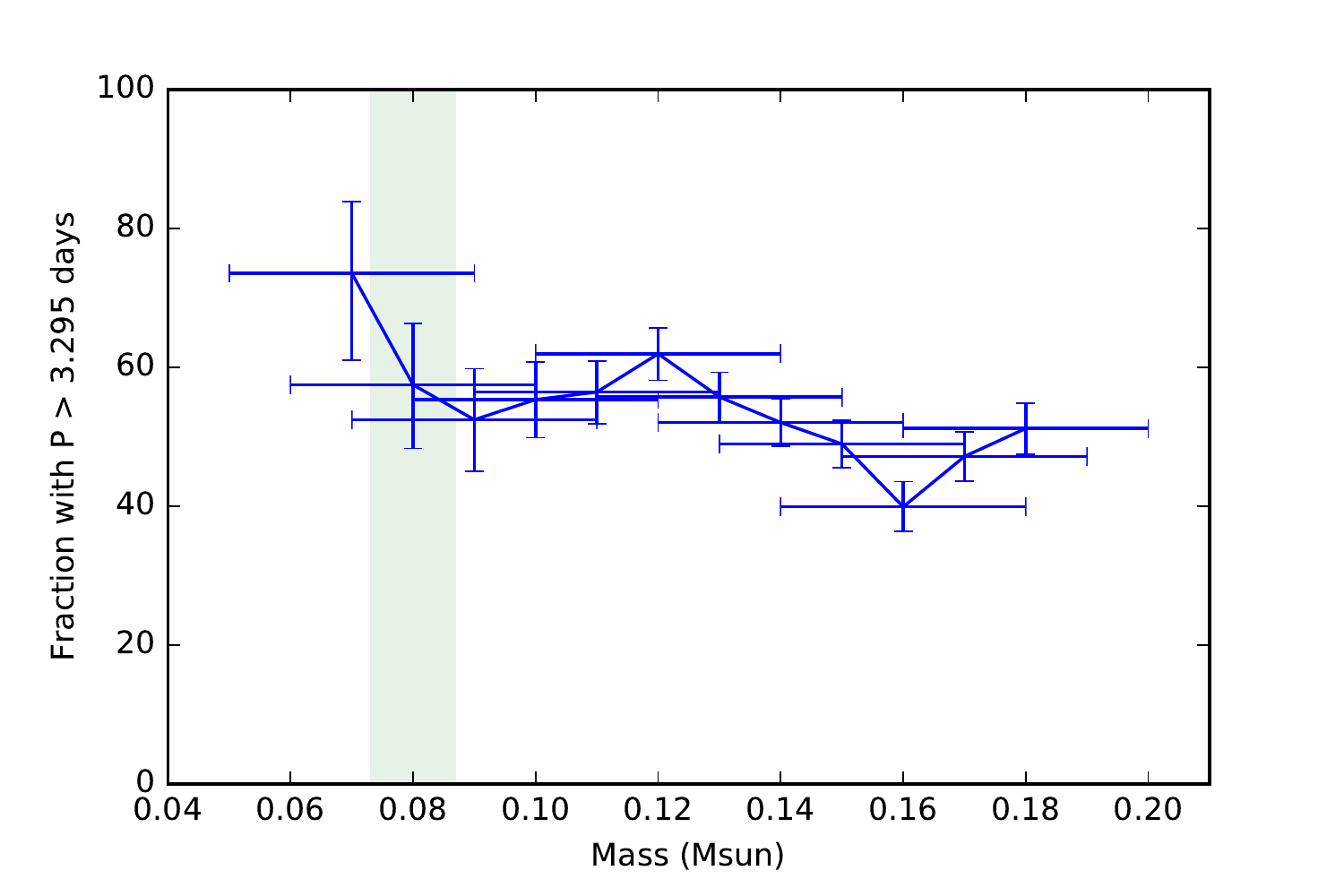}{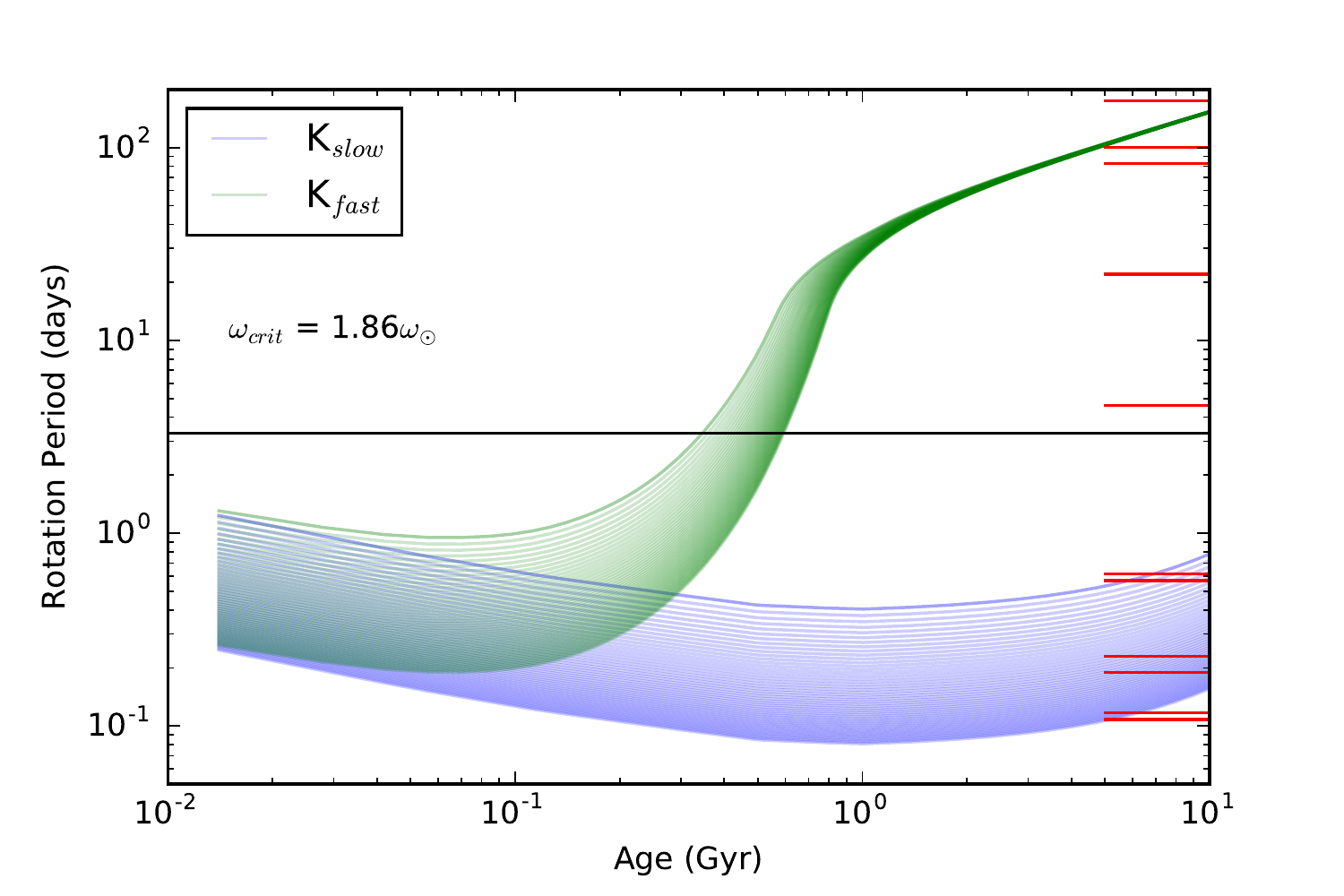}
\caption{(Left) Fraction of sources in the \citet{2016ApJ...821...93N}
  sample with rotation periods greater than 3.295~days as a function of
  mass. Horizontal error bars indicate the sample bin size, vertical
  error bars the binomial sampling uncertainties. For the mass
  estimate of TRAPPIST-1, just over half of the sample spins slower,
  suggesting this source is roughly ``middle-age'' for an ultracool M
  dwarf.  (Right) Angular momentum evolution for a 0.08~M$_{\odot}$
  star based on ``slow'' (blue lines) and ``fast'' (green lines)
  momentum loss following the prescription of
  \citet{1995ApJ...441..865C} and \citet{2011ApJ...727...56I}. The
  various lines sample initial rotation rates of
  20--100~$\omega_{\odot}$. The observed rotation period of
  TRAPPIST-1 is indicated by the solid black line; the red
  marks to the right of the panel indicate rotation periods measured
  for 0.06--0.10~M$_{\odot}$ stars by \citet{2016ApJ...821...93N}.
\label{figure:rotation} }
\end{figure}

To quantify this, we compared the rotation period of TRAPPIST-1 
to those of mid- and late-M dwarfs observed through the
MEarth program \citep{2008PASP..120..317N} as reported in
\citet{2016ApJ...821...93N}. Selecting subsamples of stars with
significant periodic variability ($A/\sigma_A$ $>$ 2) in 0.02~M$_{\odot}$ bins over masses
of 0.07--0.18~M$_{\odot}$, we computed the fraction of each subsample
that had periods longer than TRAPPIST-1. Figure~\ref{figure:rotation}
shows the resulting trend, illustrating that roughly 60\% of the stars
observed in this study were slower rotators, suggesting that they are
older. If we assume (simplistically) that rotation declines
monotonically over time, and use the GCS sample as an age prior, this
analysis would suggest an age of $\sim$4--5~Gyr for TRAPPIST-1, depending on the assumed
age of the Milky Way.

However, the rotation periods of very low mass stars at late ages 
are highly sensitive to initial conditions and the mechanism for angular momentum
loss. Figure~\ref{figure:rotation} shows the evolution of rotation
period for a 0.08~M$_{\odot}$ star following the angular
momentum loss rate prescription of \citet{1995ApJ...441..865C} and
\citet{1997ApJ...480..303K} as previously applied to low-mass stars
(e.g.,
\citealt{1997A&A...326.1023B,2008ApJ...684.1390R,2011ApJ...727...56I}). We
assumed a critical (or saturation) rotation rate $\omega_{crit}$ =
$\omega_{crit,\odot}\frac{\tau_{\odot}}{\tau}$ = 1.86$\omega_{\odot}$,
where $\omega_{crit,\odot}$ = 10$\omega_{\odot}$ and $\tau$ is the
convective overturn time assumed proportional to M$^{-2/3}$
\citep{2008ApJ...684.1390R}. We used the time-dependent radii for an
M = 0.08~M$_{\odot}$ from the models of
\citet{2003A&A...402..701B}. We considered both ``slow'' and ``fast''
prescriptions for spindown\footnote{In the notation of
  \citet{1995ApJ...441..865C} these are $K_{slow}$ =
  1.20$\times$10$^{45}$~g~cm$^2$~s = 
  1.25$\times$10$^{-10}$ M$_{\odot}$ R$_{\odot}^2$~s and $K_{fast}$ =
  1.12$\times$10$^{47}$~g~cm$^2$~s = 1.16$\times$10$^{-8}$ M$_{\odot}$ R$_{\odot}^2$~s, respectively.}
from \citet{2011ApJ...727...56I}, as well as a range of initial
rotation velocities at $\sim$20-50~Myr of 20--100~$\omega_{\odot}$,
based on measurements for M $<$ 0.35~M$_{\odot}$ stars in NGC 2547 
reported in the same study.  As shown in Figure~\ref{figure:rotation},
different angular momentum loss prescriptions produce widely dispersed 
rotation periods beyond 300~Myr, spanning three orders
of magnitude by 10~Gyr. This range is
well-matched to the range of observed rotation periods for
significantly variable 0.06--0.10~M$_{\odot}$ stars in the sample of
\citet{2016ApJ...821...93N}, which span 0.11--364~days. TRAPPIST-1's
rotation period resides between the slow and fast evolutionary tracks.
Since neither the specific mechanism of spin-down nor the initial
rotation rate are known for this source, at best we can conclude that
TRAPPIST-1 is likely older than 300~Myr, the age at which the ``fast''
track periods exceed 3.3~days.

\subsection{Age Constraints from Activity}

Low-mass stars show clear age-activity correlations related to the spindown of stars and reduction of rotationally-driven magnetic dynamos \citep{1972ApJ...171..565S,2004ApJ...614..267F,2008ApJS..178..339C}.  
As spindown timescales increase for the lowest-mass stars, saturated magnetic emission can persist for even slowly rotating stars ($P < 86$~day) with little correlation between the incidence of emission and rotation period \citep{2008AJ....135..785W,2015ApJ...812....3W}. However, there is evidence for a correlation between the strength of H$\alpha$ emission
and ``stratigraphic'' age (distance from the Galactic plane) that continues through the end of the M dwarf sequence.

\begin{figure}
\plottwo{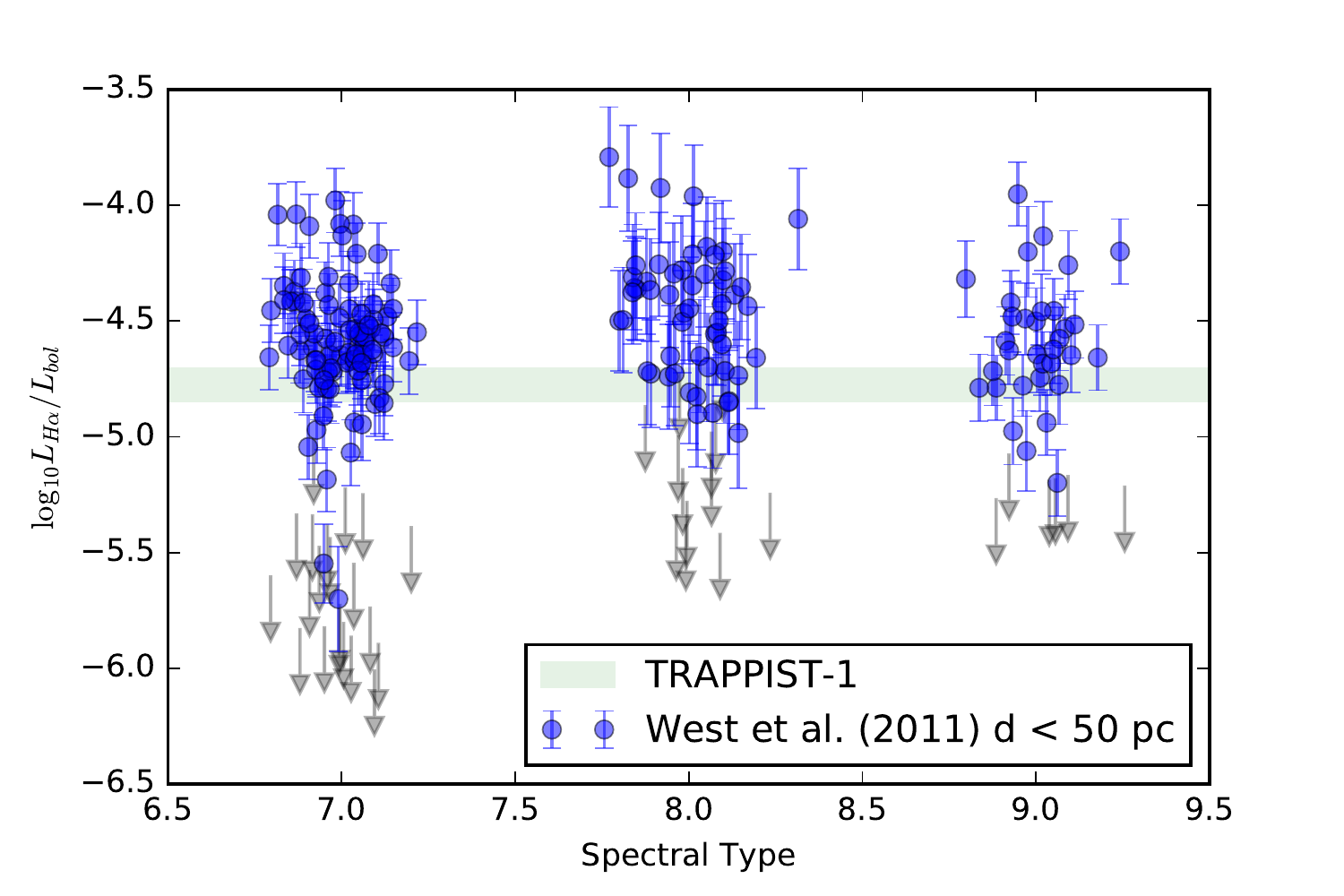}{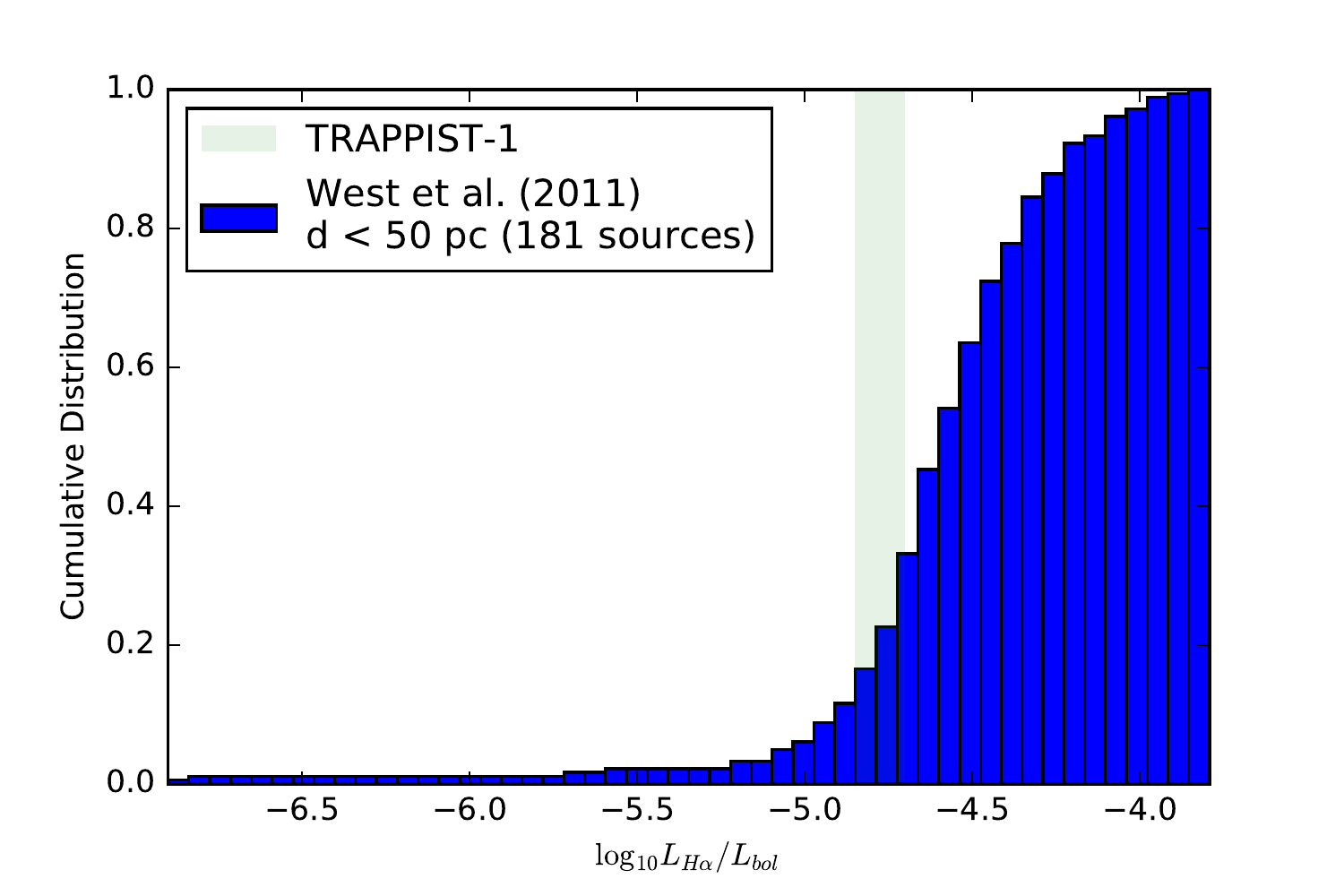}
\caption{(Left) Measurements of {\lhalbol} for 239 M7-M9 dwarfs with 50 pc with SDSS spectra as reported by 
\citet[circles and arrows]{2011AJ....141...97W} compared to the range of values reported for 
TRAPPIST-1 (green region; Table~1). Circles with error bars are sources with significant detections (H$\alpha$ emission peak more than three times the continuum noise); downward arrows are upper limits. Points are randomly offset along the x-axis to aid in visualization.
(Right) Cumulative distribution of significant {\lhalbol} measurements for the SDSS sample, again compared to the range of measurements for TRAPPIST-1. This star is less active than over half of the nearby M7-M9 dwarfs. 
\label{fig:activity} }
\end{figure}

Persistent H$\alpha$ emission from TRAPPIST-1
is weaker than emission in over half of the active late-M dwarfs near the Sun (Figure~\ref{fig:activity}),
suggesting an age in the upper half of this sample. This comparison stands in contrast to
claims that TRAPPIST-1 is ``highly active'', suggesting youth (e.g.,
\citealt{2017A&A...599L...3B,2017arXiv170310130V}).  
The perception that TRAPPIST-1 is highly active is also related to its flaring emission, and specifically the 
detection of a super-solar flare in its $K2$ lightcurve,
with an integrated energy E = 10$^{33}$ erg 
\citep{2017arXiv170310130V}.  
Again, context is essential.
M dwarfs typically have higher optical and infrared flare rates and flare
energies than solar-type stars \citep{2012ApJ...748...58D} and
TRAPPIST-1's flare duty cycle during the {\em K2} monitoring period,
$\approx$0.1\%\footnote{Based on a typical flare time scale of 1-2~min
  = 0.02--0.03~hr and the median time between flares 28.1~hr.} as
reported by \citet{2017arXiv170310130V}, is more than an order of
magnitude below the 3$\pm$1\% inferred for M7-M9 dwarfs by
\citet{2010AJ....140.1402H}.  The cumulative flare frequency
distribution for TRAPPIST-1 as a function of energy, also reported in
\citet{2017arXiv170310130V}, is similarly depressed by a factor of
$\approx$4 compared to other M6-M8 dwarfs
\citep{2011PhDT.......144H,2017arXiv170308745G}.  Finally, the {\em
  K2} flare, while dramatic, is not unique;
\citet{2017ApJ...838...22G} report an E $>$ 4$\times$10$^{33}$~erg
flare from an L0 dwarf observed with {\em K2} and
\citet{2014ApJ...781L..24S,2016ApJ...828L..22S} have reported E $>$
10$^{34}$~erg flares from M8 and L1 dwarfs detected in the All-Sky
Automated Survey for Supernovae survey \citep{2014ApJ...788...48S}.


Taken together, these activity metrics suggest TRAPPIST-1 is older than the typical late-M dwarf in the Solar Neighborhood,
but remains an active star. Given the lack of empirical calibrations for age/activity among the latest M dwarfs, we
are unable to more specifically quantify TRAPPIST-1's age from these measures. 

\section{Discussion}

\begin{figure}
\plotone{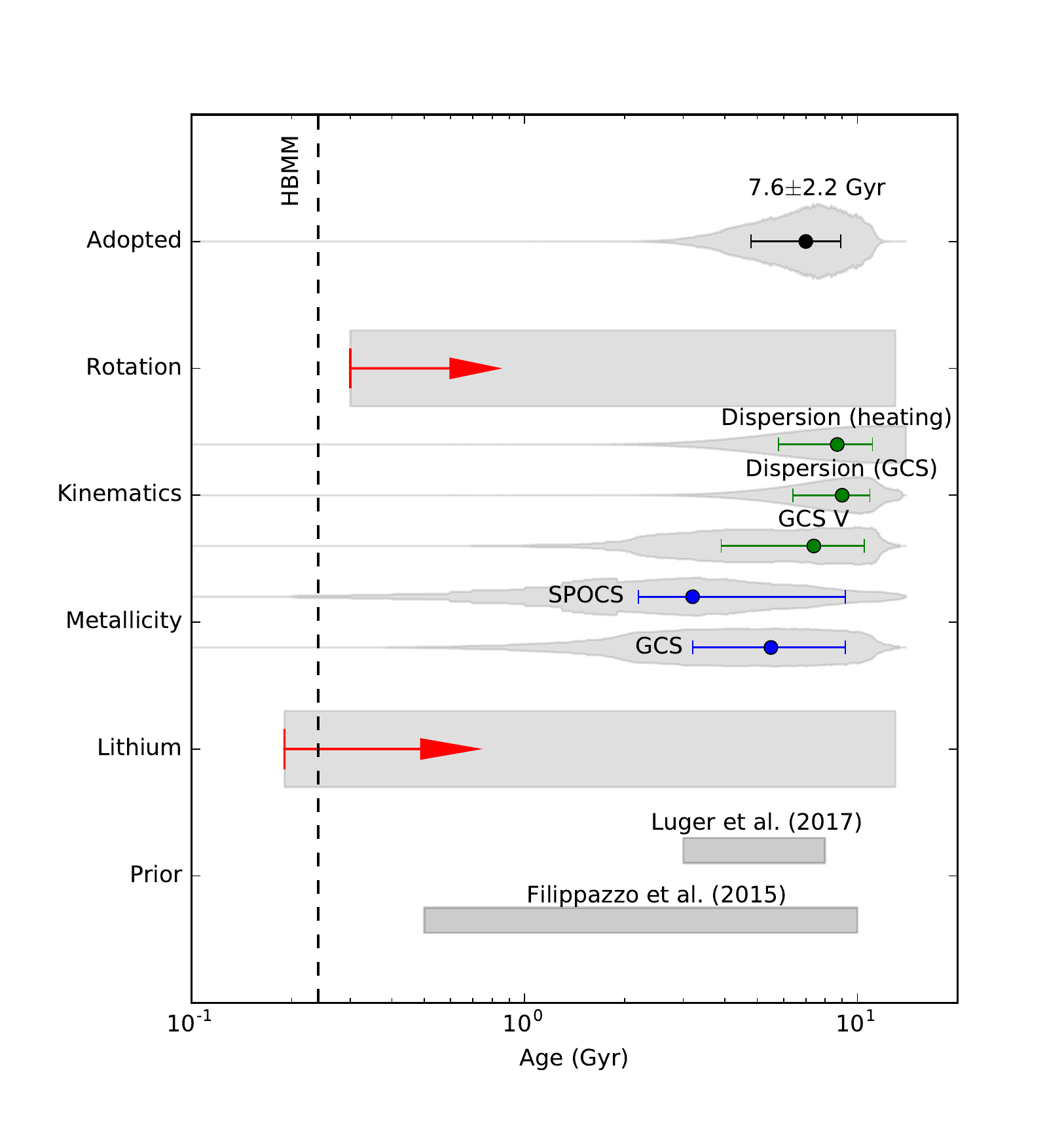}
\caption{Summary of age estimates for TRAPPIST-1, from bottom to top:
  original age range from \citet{2015ApJ...810..158F}; lower limit
  based on absence of Li~I absorption; age probability distribution
  functions for GCS and SPOCS stars with similar metallicities; age
  probability distribution function for GCS stars with $V$ $\leq$
  $V_{T1}$ = $-$66~{\kms}; kinematic dispersion simulations with age priors based on
  the GCS sample (lower) and heating losses (upper); lower limit based on rotation period; and
  our concordance age estimate. Throughout, symbols with error bars
  indicate the maximum likelihood and 16--84\% probability ranges,
  while the grey shaded regions map the underlying probability
  distribution functions.
\label{figure:summary} }
\end{figure}

Combining our age probability distribution functions from metallicity
and kinematics, and lower limits from the absence of lithium
absorption and measured rotation period, we deduce a concordance age of {\age}~Gyr for
TRAPPIST-1 (Figure~\ref{figure:summary}). This is inconsistent with 
some of the qualitative estimates reported in the literature 
(e.g., ``relatively young''; \citealt{2017A&A...599L...3B}; ``young'', \citealt{2017MNRAS.469L..26O}),
and is on the high end of the 3--8~Gyr age adopted by \citet{2017arXiv170304166L}.
This older age has important implications on both the stability and habitability of its orbiting planets.

In terms of stability, N-body simulations presented in \citet{2017Natur.542..456G} showed the planetary
system to be consistently
unstable on timescales $<$0.5~Myr, with only an 8\% change of surviving 1~Gyr.
This is refuted by the much older age we infer for the TRAPPIST-1 star.
However, recent simulations show that the resonant configuration of these
planets is in fact highly stable through disk migration on timescales of 50~Myr (10$^{10}$ orbits),
with or without eccentricity dampening. That this system appears to have persisted for over 5~Gyr, despite dynamial interactions that are readily detectable through transit timing variations
\citep{2017Natur.542..456G,2017arXiv170404290W}, suggests that the resonant configuration
is indeed inherently stable.

In terms of habitability, despite TRAPPIST-1's modest emission as compared to other
late-M dwarfs, the radiation and particle environment is still extreme as compared to the 
Earth \citep{2017MNRAS.464.3728B,2017MNRAS.465L..74W,2017reph.conf10106G}. 
Based on current estimates of XUV-driven mass loss, 
the high energy emission of TRAPPIST-1 is likely sufficient to have evaporated
an Earth's ocean of water mass from each of the TRAPPIST-1 
planets except $g$ and $h$ over the system's lifetime \citep{2017MNRAS.464.3728B,2017A&A...599L...3B}.
Moreover, the stripping of atmospheres and oceans may be enhanced by direct interaction between stellar and planetary magnetic field lines, which could funnel stellar wind particles directly to the planets' surfaces \citep{2016ApJ...833L...4G,2017reph.conf10106G}.
On the other hand, current estimates of the planets' densities are generally below Earth's average density 
\citep{2017Natur.542..456G,2017arXiv170404290W}, suggesting volatile-rich worlds that may have
ample reservoirs; while
ocean evaporation and hydrogen loss could result in an oxygen- and ozone-rich atmosphere that could shield the surface from high UV fluxes \citep{2015AsBio..15..119L,2017MNRAS.469L..26O}.
Transit spectroscopy measurements of the atmospheres of these planets are currently insufficient
to detect the signatures of all but the lightest elements \citep{2016Natur.537...69D}, but the {\it James Webb Space Telescope} should have the sensitivity to detect Earth-like atmospheres around 
these plants, if they exist \citep{2016MNRAS.461L..92B}. 

Finally, we note that agreement between the observed luminosity, average stellar density and evolutionary models can be achieved if the star's radius is modestly inflated relative to model predictions. Our analysis indicates that a radius of 0.121$\pm$0.003~R$_{\odot}$ is needed to bring both \citet{1997ApJ...491..856B,2001RvMP...73..719B} and \citet{2015A&A...577A..42B} evolutionary models in line with the observed properties of TRAPPIST-1. This radius is formally consistent with the value adopted in \citet{2016Natur.533..221G}, and represents a modest 3\% increase in planetary radii and 11\% decrease in inferred planetary densities, 
less than current uncertainties \citep{2017Natur.542..456G,2017arXiv170404290W}.

\acknowledgments

The authors acknolwedge discussions with Eric Agol, Vincent Bourrier, Amaury Triaud, and Valerie van Grootel
that aided in the preparation of the manuscript;
and thank the Hon.\ John Culberson of Texas's 7th
congressional district, US House of Representatives, for asking about
the age of TRAPPIST-1 during his visit to JPL in February 2017, which
spurred the writing of this paper.
EEM acknowledges the NASA NExSS program for support.
AJB acknowledges funding support from
the National Science Foundation under award No.\ AST-1517177.
Part of this research was carried out at the Jet Propulsion
Laboratory, California Institute of Technology, under a contract with
the National Aeronautics and Space Administration.
This document does not contain export controlled information
(URS266250).
This material is based upon
work supported by the National Aeronautics and Space
Administration under Grant No.\ NNX16AF47G issued
through the Astrophysics Data Analysis Program
This research has made use of the SIMBAD
database, operated at CDS, Strasbourg, France; NASA's
Astrophysics Data System Bibliographic Services; the M, L,
T, and Y dwarf compendium housed at \url{http://DwarfArchives.org}; 
and the Spex Prism Libraries at \url{http://www.browndwarfs.org/spexprism}.
\\

\vspace{5mm}
\software{astropy \citep{2013A&A...558A..33A},
	SPLAT \citep{2014ASInC..11....7B}}


\end{document}